\documentclass[referee]{aa} 

\usepackage{xcolor}
\usepackage{graphicx} 
\usepackage{mathtools}
\usepackage{txfonts}

\newcommand\Bv{{\bf B}}
\newcommand\Ev{{\bf E}}

\newcommand\uv{{\bf u}}

\newcommand\vv{{\bf v}}

\newcommand\Nv{{\mathbf{N}}}

\newcommand\jv{{\bf j}}

\begin{document}
 
 \title{Transition to kinetic turbulence at proton scales driven by large-amplitude Kinetic Alfv\'en fluctuations}

 \author{F. Valentini\inst{1} 
         \and C. L. V\'asconez \inst{2,3,1} 
         \and O. Pezzi \inst{1}
         \and S. Servidio \inst{1}
         \and F. Malara \inst{1}
         \and F. Pucci \inst{4}
         }
 \institute{Dipartimento di Fisica, Universit\`a della Calabria, 87036, Rende (CS), Italy 
            \and Departamento de F\'isica, Facultad de Ciencias, Escuela Polit\'ecnica Nacional, Quito, Ecuador
            \and Observatorio Astron\'omico de Quito, Escuela Polit\'ecnica Nacional, Quito, Ecuador
            \and Center for Mathematical Plasma Astophysics, Departement Wiskunde, Universiteit Leuven, Leuven, Belgium 
            }

 \date{Received...}

 \abstract{ Space plasmas are dominated by the presence of large-amplitude waves, large-scale inhomogeneities, kinetic effects and
turbulence. Beside the homogeneous turbulence, generation of small scale fluctuations can take place also in other  realistic
configurations, namely, when perturbations superpose to an inhomogeneous background magnetic field. When an Alfv\'en wave
propagates in a medium where the Alfv\'en speed varies in a direction transverse to the mean field, it undergoes phase-mixing,
which progressively bends wavefronts, generating small scales in the transverse direction. As soon as transverse scales get of the
order of the proton inertial length $d_p$, kinetic Alfv\'en waves (KAWs) are naturally generated. KAWs belong to the branch of
Alfv\'en waves and propagate nearly perpendicular to the ambient magnetic field, at scales close to $d_p$. Many numerical,
observational and theoretical works have suggested that these fluctuations may play a determinant role in the development of the
solar-wind turbulent cascade. In the present paper, the generation of large amplitude KAW fluctuations in inhomogeneous background
and their effect on the protons have been investigated by means of hybrid Vlasov-Maxwell direct numerical simulations. Imposing a
pressure balanced magnetic shear, the kinetic dynamics of protons has been investigated by varying both the magnetic configuration
and the amplitude of the initial perturbations. Of interest here is the transition from quasi-linear to turbulent regimes,
focusing, in particular, on the development of important non-Maxwellian features in the proton distribution function driven by KAW
fluctuations. Several indicators to quantify the deviations of the protons from thermodynamic equilibrium are presented. These
numerical results might help to explain the complex dynamics of inhomogeneous and turbulent astrophysical plasmas, such as the
heliospheric current sheet, the magnetospheric boundary layer and the solar corona.}


 \titlerunning{Kinetic Alfv\'en waves turbulence}
 \maketitle

\section{Introduction}
Alfv\'enic fluctuations, characterized by high velocity-magnetic field correlation and by a low level of density and magnetic
field intensity relative variations, are commonly observed in space plasmas. Starting from the pioneering work by \citet{belch71},
in-situ measurements in the solar wind have shown that Alfv\'enic fluctuations represent the main component of turbulence in
high-speed streams, at scales larger than the proton inertial length $d_p=V_{_A}/\Omega_{cp}$ [see \citet{bruno13} for a review],
$V_{_A}$ and $\Omega_{cp}$ being the Alfv\'en speed and the proton cyclotron frequency, respectively. Moreover, in recent years
the presence of velocity fluctuations propagating along the magnetic field at a speed compatible with the local Alfv\'en velocity
has also been ascertained in the solar corona \citep{tomc07,tomc09} and interpreted as Alfv\'en waves. Such waves could possibly
represent a source for Alfv\'enic fluctuations detected in the turbulence of solar wind, which emanates from the corona.

At scales comparable with $d_p$, a variety of observations in the solar wind have suggested that fluctuations may consist
primarily of Kinetic Alfv\'en waves (KAWs) \citep{bale05,sahraoui09}. In the linear fluctuation terminology, KAWs are waves 
belonging to the Alfv\'en branch, at wavevectors ${\bf k}$ nearly perpendicular to the ambient magnetic field ${\bf B}_0$, with $k
\sim d_p^{-1}$. A detailed discussion of the properties of KAWs can be recovered, e.g., in \citet{hollweg99} (see also references 
therein for a more complete view of the subject). In the last decades KAWs have received a considerable attention due to their
possible role in a normal mode description of turbulence. Indeed, theoretical studies (e.g.,
\citet{shebalin82,carbone90,OughtonEA94}) have shown that the turbulent cascade in a magnetized plasma tends to develop mainly in
the directions perpendicular to ${\bf B}_0$. Anisotropic spectra have been commonly observed in space plasmas, showing the
presence of a significant population of quasi-perpedicular wavevectors \citep{matthaeus86,matthaeus90}. The above considerations
suggest that fluctuations with characteristics similar to KAWs are naturally generated by a turbulent cascade at scales $\sim
d_p$. Many solar wind observational studies \citep{bale05,sahraoui09,podesta12,salem12,chen13,kiyani13}, theoretical works
\citep{howes08a,scheko09,sahraoui12} as well as numerical simulations \citep{gary04,howes08b,tenbarge12} have suggested that
fluctuations near the end of the magnetohydrodynamics inertial cascade range may consist primarily of KAWs, and that such
fluctuations can play an important role in the dissipation of turbulent energy. Due to a nonvanishing parallel component of the
electric field associated with KAWs, these waves have also been considered in the problem of particle acceleration
\citep{voitenko04,decamp06}. Particle acceleration in Alfv\'en waves in a dispersive regime has been studied both in 2D
\citep{tsiklauri05,tsiklauri11} and in 3D \citep{tsiklauri12} configurations. Recently, \citet{vasconez14} have studied
collisionless Landau damping and wave-particle resonant interactions in KAWs.

Beside the homogeneous turbulence, generation of small scale fluctuations takes place also in more realistic configurations,
namely, when perturbations superpose to an inhomogeneous mean field ${\bf B}_0(\bf x)$. For instance, an Alfv\'en wave
propagating in a medium where the Alfv\'en velocity varies in a direction transverse to ${\bf B}_0$ undergoes phase-mixing
\citep{heyvaerts83}, which progressively bends wavefronts thus generating small scales in the transverse direction. Linear wave
propagation in a transverse-structured background and the resulting production of small scales in the magnetohydrodynamic (MHD)
regime has been extensively studied both analytically and numerically
\citep{mok85,steinolfson85,lee86,davila87,hollweg87,califano90,califano92,malara92,malara96,nakariakov97,kaghashvili99,
tsiklauri02,tsiklauri03,ofman10,ozak15}. Alfv\'en waves propagating on equilibria containing an X-type magnetic null point
\citep{landi05,mclaughlin11,pucci14}, as well as in 3D configurations in the WKB limit
\citep{similon89,petkaki98,malara00,malara03,malara05,malara07}, have also been considered, finding a fast formation of
small-scale structures transverse to the background magnetic field. Similar ideas involving dissipative mechanisms related to
interaction of Alfv\'en waves or KAWs and phase mixing have been examined in the context of the magnetospheric plasma sheet
\citep{LysakSong11} and in coronal loops \citep{OfmanAschwanden02}. In all those configurations small scales are formed in
consequence of the coupling between the wavevector ${\bf k}_0$ associated with the background inhomogeneity and the wavevector
${\bf k}$ associated with the perturbation. However, this effect appears also in the context of nonlinear MHD when imposed
parallel propagating waves interact with an inhomogeneous background consisting either of pressure balanced structures or
velocity shears \citep{GhoshEA98}.

The above considerations indicate that, when Alfv\'en waves propagate in a background which is inhomogeneous in the direction
transverse to ${\bf B}_0$, KAWs should naturally form as soon as transverse wavevectors of the order of $d_p^{-1}$ are generated
by the wave-inhomogeneity coupling. This mechanism has been investigated in recent works \citep{vasconez15,pucci16}, where the
evolution of an initial Alfv\'en wave with different polarizations propagating in a pressure-balanced inhomogeneous equilibrium
has been analyzed numerically. These studies have been carried out by means of both Hall-MHD and hybrid Vlasov-Maxwell (HVM)
simulations. The former include dispersive effects determining the evolution of structures at $k\sim d_p$, with a limited
computational effort; the latter allow for a description of kinetic effects related to the evolution of the proton velocity
distribution function (VDF). The results have shown that in all the considered configurations the time evolution of initially
linearly polarized Alfv\'en waves leads to the generation of KAWs in the inhomogeneity regions of the equilibrium structure. This
happens both in cases  when phase-mixing is active and when it is absent. Moreover, HVM simulations carried out for waves with
moderate amplitudes have shown the presence of kinetic features related to departure from maxwellianity of the proton VDF with
(i) $T_\perp \neq T_{||}$, and (ii) the presence of beams of protons accelerated along the background magnetic field
\citep{valentini11} at a speed comparable with the parallel phase velocity of the waves. Both features (i) and (ii) are spatially
localized at KAWs locations; moreover, the presence of proton beams is probably related to a parallel electric field component
associated with KAWs. The above quasi-linear studies suggest that the dynamics of Alfv\'en waves with shears can be crucial for
the understanding of more complex (and realistic) scenarios, such as the turbulent solar wind, the magnetosheet and the
inhomogeneous regions of the solar corona. Hence the keypoint is now to understand the transition from KAWs to turbulence.

In the present paper we use 2D-3V HVM simulations (two dimensions in physical space and three dimensions in velocity space)
to investigate the dynamics of Alfv\'en waves with inhomogeneous magnetic configurations, at scales comparable with the proton
skin depth, varying both the background equilibrium and the fluctuations amplitude. Previous HVM simulations
\citep{servidio12,servidio14,servidio15} carried out for more turbulent configurations have shown the formation of local
departures from maxwellianity in the proton VDF, such as temperature anisotropy, with a significant dependence on the value of
the proton plasma $\beta$ (ratio between kinetic and magnetic pressure). Here, we consider both moderate and large-amplitude
perturbations, hence approaching to turbulence, giving a quantitative characterization of the modifications the proton VDF
undergoes to, in consequence of interactions with the perturbations.

The outline of the paper is as follows. In Section 2, we present the HVM equations and the setup of the numerical runs. In
Section 3, we discuss the results of two HVM simulations obtained when varying the parameters of the initial equilibrium and the
amplitude of the initial perturbations, focusing, in particular, on the quantification of the nonlinear departures of the proton
VDF from local Maxwellian equilibrium. We finally conclude and summarize in Section 4.

\section{Hybrid Vlasov-Maxwell simulation setup}
We solve numerically the HVM equations \citep{valentini07} in 2D-3V phase space configuration. The equations of the HVM system in
physical units are written as:
\begin{eqnarray}
 \label{eq:vlasovph}
& & \frac{\partial f}{\partial t} + \vv \cdot {\nabla} f+ \frac{e}{m_p}\left( \Ev + 
\frac{1}{c}\vv \times \Bv \right)\cdot\frac{\partial f}{\partial \vv}
= 0\\
\label{eq:ohm3ph}
& &{\bf E}  = - \frac{1}{c}({\bf u} \times {\bf B}) + 
\frac{1}{en}\left( \frac{{\bf j} \times {\bf B}}{c}\right) -\frac{1}{en} \nabla P_e\\ 
\label{eq:Maxw_bph}
& &\frac{\partial {\Bv}}{\partial t} = - c \nabla \times {\Ev};\;\;\;
{\jv} = \frac{c}{4\pi}\nabla \times {\Bv} 
\end{eqnarray}
\noindent
where $f$ is the proton distribution function, $\Ev$ and $\Bv$ the electric and magnetic fields, respectively, ${\bf j}$ the total
current density (the displacement current has been neglected in the Ampere equation and quasi-neutrality is assumed), $m_p$ and
$e$ are the proton mass and charge, respectively, and $c$ is the velocity of light. The proton density $n$ (which is equal to the
electron density) and bulk velocity ${\uv}$ are obtained as velocity moments of $f$. The scalar electron pressure $P_e$ is
assigned through an isothermal equation of state $P_e = \kappa_B n T_e$, where $T_e={\rm const}$ is the electron temperature and
$\kappa_B$ the Boltzmann constant. The above equations can be re-written in a dimensionless form using the following standard
procedure. We consider a typical value ${\tilde B}$ of the magnetic field and a typical density ${\tilde n}$. Using these
quantities we build up a typical Alfv\'en velocity ${\tilde c}_A = {\tilde B}/(4\pi m_p {\tilde n})^{1/2}$, a gyration frequency 
${\tilde \Omega}_p=e {\tilde B}/(m_p c)$, a typical proton inertial length ${\tilde d}_p={\tilde c}_A/{\tilde \Omega}_p$, a
typical current density ${\tilde j}=c {\tilde B}/(4\pi {\tilde d}_p)$, and a typical temperature ${\tilde T}={\tilde B}^2/(4\pi
\kappa_B {\tilde n})$. Then, we normalize the electric and magnetic fields ${\bf B}$ and ${\bf E}$ to ${\tilde B}$; the velocities
${\bf v}$ and ${\bf u}$ to ${\tilde c}_A$; the density $n$ to ${\tilde n}$; the current density ${\bf j}$ to ${\tilde j}$; the
electron temperature $T_e$ to ${\tilde T}$; the space variables to ${\tilde d}_p$ and time to ${\tilde \Omega}_p^{-1}$. The
equations (\ref{eq:vlasovph})-(\ref{eq:Maxw_bph}) are written in terms of the above defined dimensionless quantities in the
following form:
\begin{eqnarray}
 \label{eq:vlasov}
& & \frac{\partial f}{\partial t} + \vv \cdot {\nabla} f+ \left( \Ev + \vv \times \Bv \right)\cdot\frac{\partial f}{\partial \vv}
= 0\\
\label{eq:ohm3}
& &{\bf E}  = - ({\bf u} \times {\bf B}) + \frac{1}{n}({\bf j} \times {\bf B}) -\frac 1 n \nabla P_e\\ 
\label{eq:Maxw_b}
& &\frac{\partial {\Bv}}{\partial t} = - \nabla \times {\Ev};\;\;\;\nabla \times {\Bv} = {\jv}
\end{eqnarray}
\noindent
where, for simplicity, each dimensionless quantity is indicated using the same notation as the corresponding physical quantity. In
what follows, all the results will be expressed in terms of the above-defined dimensionless quantities.

The normalized HVM equations (\ref{eq:vlasov})-(\ref{eq:Maxw_b}) have been solved in a double periodic spatial domain $D = L
\times L = [0, 16\pi] \times [0,16 \pi]$. In the three-dimensional velocity box, the distribution function $f$ is set equal to
zero for $|v|>v_{max} = 5 v_{thp}$ in each velocity direction. Physical space is discretized with $N_x=256$ grid points in the $x$
direction and $N_y=1024$ gridpoints in the $y$ direction, while velocity space with $N_{V_x} =N_{V_y} = N_{V_z} = 51$ grid points.

When dealing with nonuniform situations in kinetic regime, the definition of an equilibrium state is a delicate point
\citep{cerri13,cerri14}. Here, we consider a nonuniform equilibrium configuration, where physical quantities vary only along the
$y$-direction, or are uniform. Quantities relative to the equilibrium state are indicated by the upper index ``(0)". The
equilibrium magnetic field is directed along the $x$-direction:
\begin{equation}\label{B0}
{\bf B}^{(0)}=B^{(0)}(y)\, {\bf e}_x
\end{equation}
where ${\bf e}_x$ is the unit vector in the $x$ direction, while the equilibrium proton distribution function is a Maxwellian with
the following form:
\begin{equation}\label{distr0}
f^{(0)}(y,{\bf v})=\frac{n^{(0)}(y)}{(2\pi T_p^{(0)})^{3/2}} \exp \left( 
-\frac{v_x^2+v_y^2+v_z^2}{2T_p^{(0)}}\right)
\end{equation}
where the function $n^{(0)}(y)$ represents the nonuniform density:
\begin{equation}\label{n0}
\int f^{(0)}(y,{\bf v}) \, d^3 {\bf v} = n^{(0)}(y)
\end{equation}
and $T_p^{(0)}$ is the equilibrium proton temperature which we assume to be uniform. Moreover, the corresponding equilibrium
proton bulk velocity is vanishing:
\begin{equation}\label{u0}
{\bf u}^{(0)}(y)=\frac{1}{n^{(0)}(y)}\int {\bf v} f^{(0)}(y,{\bf v})\, d^3 {\bf v} = 0
\end{equation}
From equation (\ref{eq:ohm3}) and using equations (\ref{B0}) and (\ref{u0}), the equilibrium electric field can be derived:
\begin{equation}\label{E0}
{\bf E}^{(0)}=-\frac{1}{n^{(0)}(y)} \left\{ \left[ \frac{d}{d y}
\left( \frac{{B^{(0)}}^2}{2} \right)\right] + \frac{d n^{(0)}}{dy} T_e \right\}
{\bf e}_y
\end{equation}
where ${\bf e}_y$ is the unit vector in the $y$ direction. We note that $\nabla \times {\bf E}^{(0)}=0$; thus, according to the
Faraday law (\ref{eq:Maxw_b}), the magnetic field ${\bf B}^{(0)}$ remains constant in time. Finally, we consider the Vlasov
equation, calculating the single terms in equation (\ref{eq:vlasov}):
\begin{equation}\label{term1}
{\bf v} \cdot \nabla f^{(0)} = v_y \frac{\partial f^{(0)}}{\partial y} = 
v_y \frac{dn^{(0)}}{dy} \frac{f^{(0)}}{n^{(0)}}
\end{equation}
\begin{equation}\label{term2}
{\bf E}^{(0)} \cdot \frac{\partial f^{(0)}}{\partial {\bf v}} = 
\left[ \frac{d}{dy} \left( \frac{{B^{(0)}}^2}{2} \right) + T_e \frac{d n^{(0)}}{dy}\right]
\frac{f^{(0)}}{n^{(0)}} \frac{v_y}{T_p^{(0)}}
\end{equation}
\begin{equation}\label{term3}
\left( {\bf v}\times {\bf B}^{(0)}\right) \cdot \frac{\partial f^{(0)}}{\partial {\bf v}}=
\frac{B^{(0)}\, f^{(0)}}{T_p^{(0)}} \left( v_z v_y - v_y v_z \right) = 0
\end{equation}
Using (\ref{term1})-(\ref{term3}), the equation (\ref{eq:vlasov}) gives:
\begin{equation}\label{vlasov0}
\frac{\partial f^{(0)}}{\partial t} + 
\frac{v_y f^{(0)}}{n^{(0)} T_p^{(0)}} \frac{\partial}{\partial y} \left[ n^{(0)} 
\left( T_p^{(0)} + T_e \right) + \frac{{B^{(0)}}^2}{2} \right] = 0
\end{equation}
Then, the proton distribution function $f^{(0)}$ remains stationary if the total (proton + electron + magnetic) pressure is
uniform:
\begin{equation}\label{ptotconst}
P_T^{(0)} \equiv n^{(0)} \left( T_p^{(0)} + T_e \right) + \frac{{B^{(0)}}^2}{2} = {\rm const}
\end{equation}
In conclusion, the considered configuration represent an equilibrium state, provided that the pressure balance condition
(\ref{ptotconst}) is satisfied. We note that such a state corresponds also to a MHD equilibrium: in fact, since the magnetic
tension associated to ${\bf B}^{(0)}$ is vanishing (equation (\ref{B0})), the condition (\ref{ptotconst}) expresses the
equilibrium among all the forces acting on the magnetofluid.

More specifically, for the equilibrium magnetic field we used the following form:
\begin{equation}\label{B0s}
B^{(0)}(y)=b_m+\frac{b_M -b_m}{1+\left( \displaystyle{\frac{y-8\pi}{16\pi h}}\right)^r}
+\alpha \left( \frac{y}{8\pi}-1 \right)^2
\end{equation}
which defines a magnetic field which is maximum at the center $y=8\pi$ of the domain and minimum at the borders $y=0$ and
$y=16\pi$. We used the values $r=10$ and $h=0.2$, which give a nearly homogenous field both in the central part and in the two
lateral regions of the domain; these homogeneity regions are separated by two sharp shear layers which are located 
around $y=4\pi$ and $y=12\pi$, respectively. The last term in equation (\ref{B0s}) is a small correction which has been introduced
in order to get null first derivative of $B^{(0)}(y)$ at $y=0$ and $y=16\pi$. This is obtained using the following value for the
parameter $\alpha$:
\begin{equation}\label{alpha}
\alpha=\frac{(b_M-b_m)r}{2 (2h)^r \left[ 1+\left(\displaystyle{\frac{1}{2h}}\right)^r \right]^2}
\simeq 2.62 \times 10^{-4}
\end{equation}
In this case, both $B^{(0)}(y)$ and $dB^{(0)}/dy$ are periodic functions in the interval $\lbrack 0,16\pi \rbrack$. However, higher
order derivatives of $B^{(0)}(y)$ are not exactly periodic. For this reason, the expression of $B^{(0)}$ in Eq. (\ref{B0s}) has
been corrected by filtering out harmonics with wavenumbers larger than $70$ in its spectrum. This filtering procedure does not
alter the profile $B^{(0)}(y)$. The maximum value of $B^{(0)}(y)$ is given by the parameter $b_M=B^{(0)}(y=8\pi)$, while the
minimum is $B^{(0)}(y=0)=B^{(0)}(y=16\pi)\simeq b_m$. We performed three runs (RUN I, RUN II and RUN III) with different
values of the parameters $b_M$ and $b_m$, which are given in Table \ref{tbl}. In the three runs the jump in the magnetic field
magnitude through the shear regions is the same, while in RUN I values of both $b_M$ and $b_m$ larger than in RUN II and RUN III
have been used.

The proton equilibrium temperature is assumed to be equal to the electron temperature $T_p^{(0)}=T_e=T^{(0)}$. The equilibrium
density $n^{(0)}(y)$ has been determined by imposing the total pressure equilibrium (\ref{ptotconst}):
\begin{equation}\label{presseq}
n^{(0)}(y)=\frac{1}{2 T^{(0)}} \left(P_T^{(0)} - \frac{{B^{(0)}(y)}^2}{2}\right)
\end{equation}
The values of the total pressure $P_T^{(0)}$ an of the temperature $T^{(0)}$ which have been used for the three runs are given in
Table \ref{tbl}. The condition (\ref{presseq}) ensures that the considered equilibrium is a stationary state of the system. This
has been explicitly verified by performing a run (not shown) in which the above equilibrium state is used as initial condition.

\begin{figure*}
 \centering
 \includegraphics[width=0.9\textwidth,keepaspectratio]{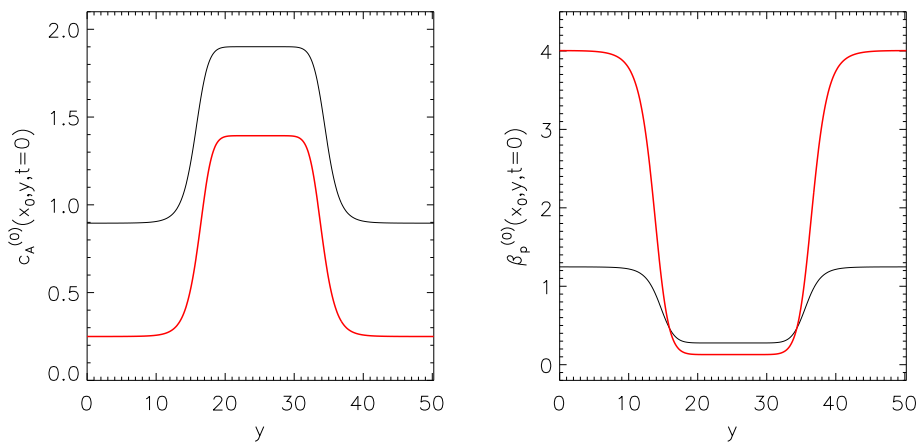}
 \caption{Profiles of $c^(0)_A (y)$ (left panel) and $\beta_p(y)$ (right panel) for RUN I (black curve) and RUN II and RUN III
(red curve).}
 \label{Fig:alfbeta}
\end{figure*}

We define a nonuniform Alfv\'en velocity associated with the equilibrium structure as
$c_A^{(0)}(y)=B^{(0)}(y)/\lbrack{n^{(0)}(y)}\rbrack^{1/2}$ and a nonuniform proton plasma beta at equilibrium as
$\beta_p^{(0)}(y)=2T^{(0)}/c_A^2$. In Figure \ref{Fig:alfbeta} the profiles of $c_A^{(0)}(y)$ (left panel) and $\beta_p^{(0)}(y)$
(right panel) are shown for RUN I (black curve) and for RUN II and RUN III (red curve): the shear layers and the homogeneity
regions are clearly visible, where $\beta_p^{(0)}$ have values either larger or smaller than the unity.

\begin{table}
\begin{center}
\caption{Simulations setup.\label{tbl}}
\begin{tabular}{c|ccccc}
\hline\hline
RUN   & $b_M$ & $b_m$ & $T^{(0)}$ & $P_T^{(0)}$ & $a$ \\
\hline\\
I  & 1.5 & 1 & 0.5 & 1.748 &  0.2 \\
\\
II & 0.8 & 0.3 & 0.125 & 0.4 &  0.2 \\
\\
III  & 0.8 & 0.3 & 0.125 & 0.4 &  0.3 \\

\hline
\end{tabular}
\end{center}
\end{table}

At the initial time $t=0$ a linearly polarized Alfv\'enic perturbation is superposed on the above-defined equilibrium. The initial
value of the magnetic field (equilibrium + perturbation) is
\begin{equation}\label{Bt0}
{\bf B}(x,y,t=0)={\bf B}^{(0)}(y) + {\bf B}^{(1)}(x) = {\bf B}^{(0)}(y) + a \cos (x/8) {\bf e}_z
\end{equation}
with ${\bf e}_z$ the unit vector in the $z$ direction, and $a$ the amplitude of the initial perturbation. The initial proton
distribution function is a shifted Maxwellian of the following form
\begin{equation}\label{ft0}
f(x,y,{\bf v},t=0) = 
\frac{n^{(0)}(y)}{(2\pi T^{(0)})^{3/2}} \exp \, \left\{ 
-\frac{v_x^2+v_y^2+\left[ v_z-u^{(1)}(x,y)\right]^2}{2T^{(0)}}\right\}
\end{equation}
where
\begin{equation}
u^{(1)}(x,y)=-\frac{a}{\left[ n^{(0)}(y)\right]^{-1/2}} \cos (x/8)
\end{equation}
It can be verified that the density and temperature associated with the initial proton distribution function (\ref{ft0}) are,
respectively:
\begin{equation}\label{n0T0}
n(x,y,t=0) = n^{(0)}(y) \;\;\; , \;\;\; T(x,y,t=0) = T^{(0)}
\end{equation}
indicating that the initial density and temperature perturbations are both vanishing. Moreover, the initial proton bulk velocity
is
\begin{equation}\label{ut0}
{\bf u}(x,y,t=0) \equiv \frac{1}{n(x,y,t=0)} \int {\bf v} f(x,y,{\bf v},t=0) \, d^3{\bf v} 
= u^{(1)}(x,y) {\bf e}_z
\end{equation}
so that ${\bf u}(x,y,t=0)=-(c_A^{(0)}/B^{(0)}){\bf B}^{(1)}$. The values of the perturbation amplitude $a$ are given in Table
\ref{tbl}.

In the three runs both the equilibrium structure and the perturbation amplitude are modified, so to increase the level of
nonlinearity when going from RUN I to RUN III. In fact, in RUN II the perturbation amplitude does not change with respect to RUN
I, but the average equilibrium magnetic field intensity is smaller than in RUN I, implying a larger ratio $a/B^{(0)}(y)$. Then, we
expect that nonlinear effects are more relevant in RUN II than in RUN I. In particular, nonlinearities are more relevant in the
lateral regions of $D$, where $B^{(0)}$ has lower values. Nonlinear effects further increase in RUN III, where the perturbation
amplitude is increased by a factor 1.5, while the same equilibrium as in RUN II is considered. Finally, we note that the profiles
of $\beta_p^{(0)}$ are different for the three runs, the configuration of RUN II and RUN III corresponding to a mean
$\beta_p^{(0)}$ larger than in RUN I.

\section{Results}
As recently discussed in \citet{vasconez15,pucci16}, when initializing the HVM simulations with the configuration described above,
the mechanism of phase-mixing of large-scale parallel propagating Alfv\'en waves in the shear regions produces KAW fluctuations at
wavelengths close to $d_p$ and at large propagation angle with respect to the magnetic field. These perturbations, while
propagating in the $x$ direction, drift along $y$ towards the boundaries of the simulation box, due to a nonvanishing transverse
component of the group velocity. Here, we present a detailed study of the role of kinetic effects on protons associated with the
propagation of KAWs produced by the above mechanism, as dependent on the characteristic of the initial equilibrium and/or on the
amplitude of the initial Alfv\'enic perturbation. In particular, we will focus on the deviations of the proton VDF from
thermodynamic equilibrium and report on the transition to a turbulent state observed when increasing the amplitude of the initial
disturbance as compared to the background values. 

\begin{figure*}
 \centering
 \includegraphics[width=0.7\textwidth,keepaspectratio]{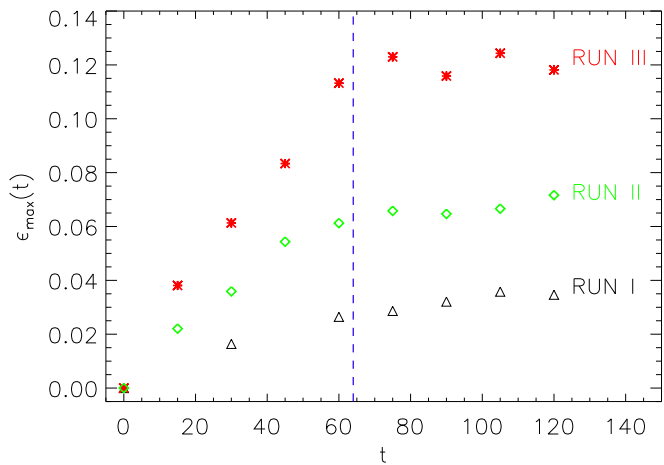}
 \caption{Time evolution of the parameter $\epsilon_{max}$ (detailed in the text) computed from the VDFs of 
RUN I (black triangles) and RUN II (red stars). 
The vertical blue-dashed line corresponds to the theoretical estimation of the time at which 
phase-mixing produces transverse scales comparable to $d_p$ .}
 \label{Fig:eps}
\end{figure*}

In order to characterize and compare the three runs, we have computed the quantity $\epsilon(x,y,t)$ \citep{greco12}, which is a
measure of the deviation of the proton VDF from the Maxwellian configuration shape:
\begin{equation}
 \epsilon(x,y,t)=\frac{1}{n({\bf x},t)} \sqrt{\int \left[f({\bf x}, {\bf v}, t) - M({\bf x}, {\bf v}, t)\right]^2d^3v}.
\end{equation}
Here $M$ is the corresponding Maxwellian distribution with the same density, bulk velocity and isotropic temperature as $f$;
$\epsilon$ is a positive definite quantity and may be viewed as a “distance” or separation between the computed $f$ and an
equivalent Maxwellian. Figure \ref{Fig:eps} shows the time evolution of $\epsilon_{max} (t)=\max_{D}\{\epsilon(x,y,t)\}$ (the
maximum value of $\epsilon$ over the two-dimensional domain $D$), from the simulations as in Table \ref{tbl}. From this figure it
can be seen that $\epsilon_{max}$ grows in time for all simulations reaching a nearly constant saturation value after the time
$t_d\simeq 60$, that gives an estimation of the typical time at which phase mixing produces transverse scales of the order of
$d_p$ \citep{vasconez15}. It is clear from this picture that the saturation level of $\epsilon_{max}$ increases as the
initial perturbation amplitude to the mean equilibrium magnetic field intensity ratio increases. In fact, going from RUN I,
through RUN II, to RUN III nonlinear effects become more and more relevant, and kinetic processes work more and more efficiently
to drive the protons away from thermodynamic equilibrium.

\begin{figure*}
 \centering
 \includegraphics[width=1.\textwidth,keepaspectratio]{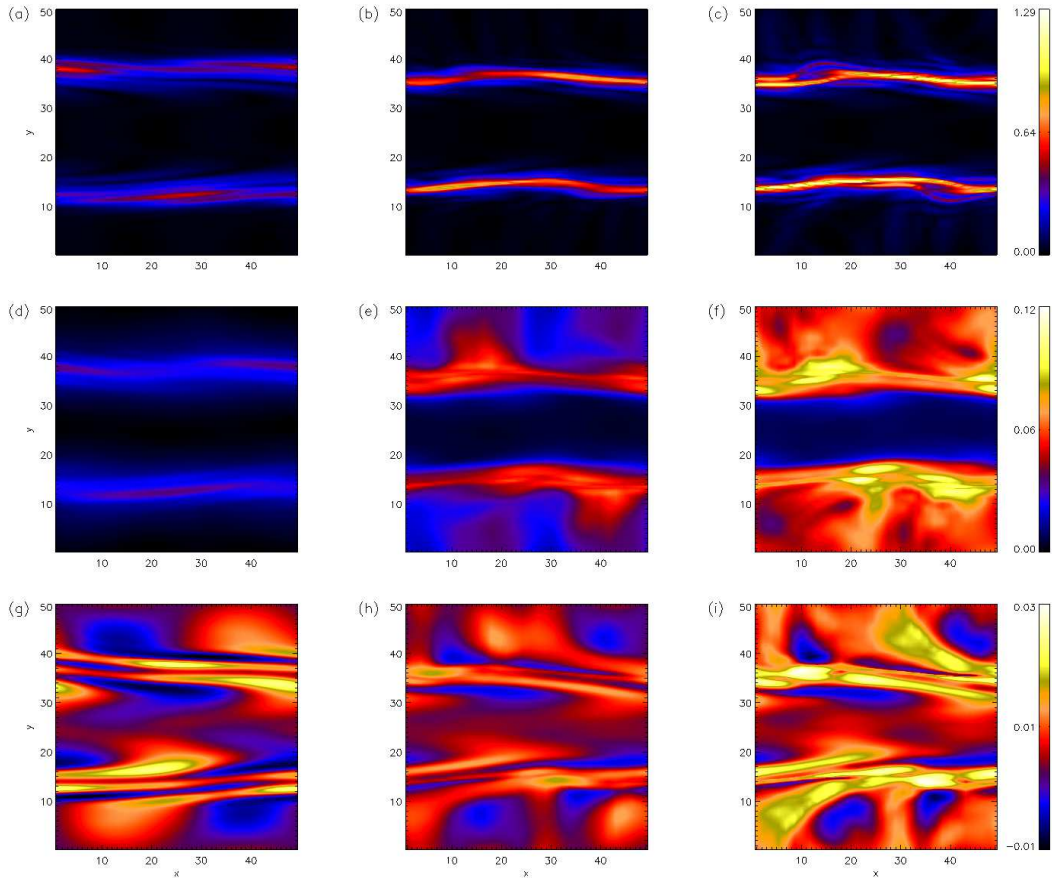}
 \caption{2D contour plots, at $t = 105$, of the modulus of the current density $| \jv |$ (upper row), 
the non-Maxwellianity measure $\epsilon$ (middle row), and temperature variations $\delta T$ (lower row), 
for RUN I (left column), RUN II (middle column), and RUN III (right column).}
 \label{Fig:contours}
\end{figure*}

The subsequent analysis of the system has been performed at a given time ($t=t^\ast=105$), at which $\epsilon_{max}$ has reached
its saturation level for all the runs and kinetic processes have significantly influenced the proton dynamics. In Figure
\ref{Fig:contours} the contour plots of $|j|$ (upper row), $\epsilon$ (middle row) and $\delta T=T-T^{(0)}$ (lower row) are
reported for RUN I (left column), RUN II (central column) and RUN III (right column). Going from RUN I to RUN III, the current
density, originally concentrated near the shear regions [panel (a)], becomes more intense and tends to filament when kinetic
physics gets dominant [panel (c)]. Non-Maxwellian features in RUN I are essentially located within the shear regions, near the
peaks of $|j|$ [panel (d)]. In RUN II, such features are still peaked in the shear regions where dispersive effects responsible
for the KAW formation are active [panel (e)]. However, significant departures from Maxwellianity are visible also in the lateral
homogeneous regions, starting from the early stage of the simulation (not shown). This becomes more evident in RUN III [panel
(f)]. These latter features can be ascribed to nonlinear effects which are particularly strong in the lateral regions, where the
ratio $a/B^{(0)}$ is of the order of unity. Finally, also temperature variations exhibit a behavior similar to that of $\epsilon$,
being more intense in the shear regions for RUN I [panel (g)] as compared to RUN II [panel (h)] and RUN III [panel (i)], in which
significant values of $\delta T$ are recovered in the whole spatial domain.

\begin{figure}
 \centering
 \includegraphics[width=0.55\textwidth]{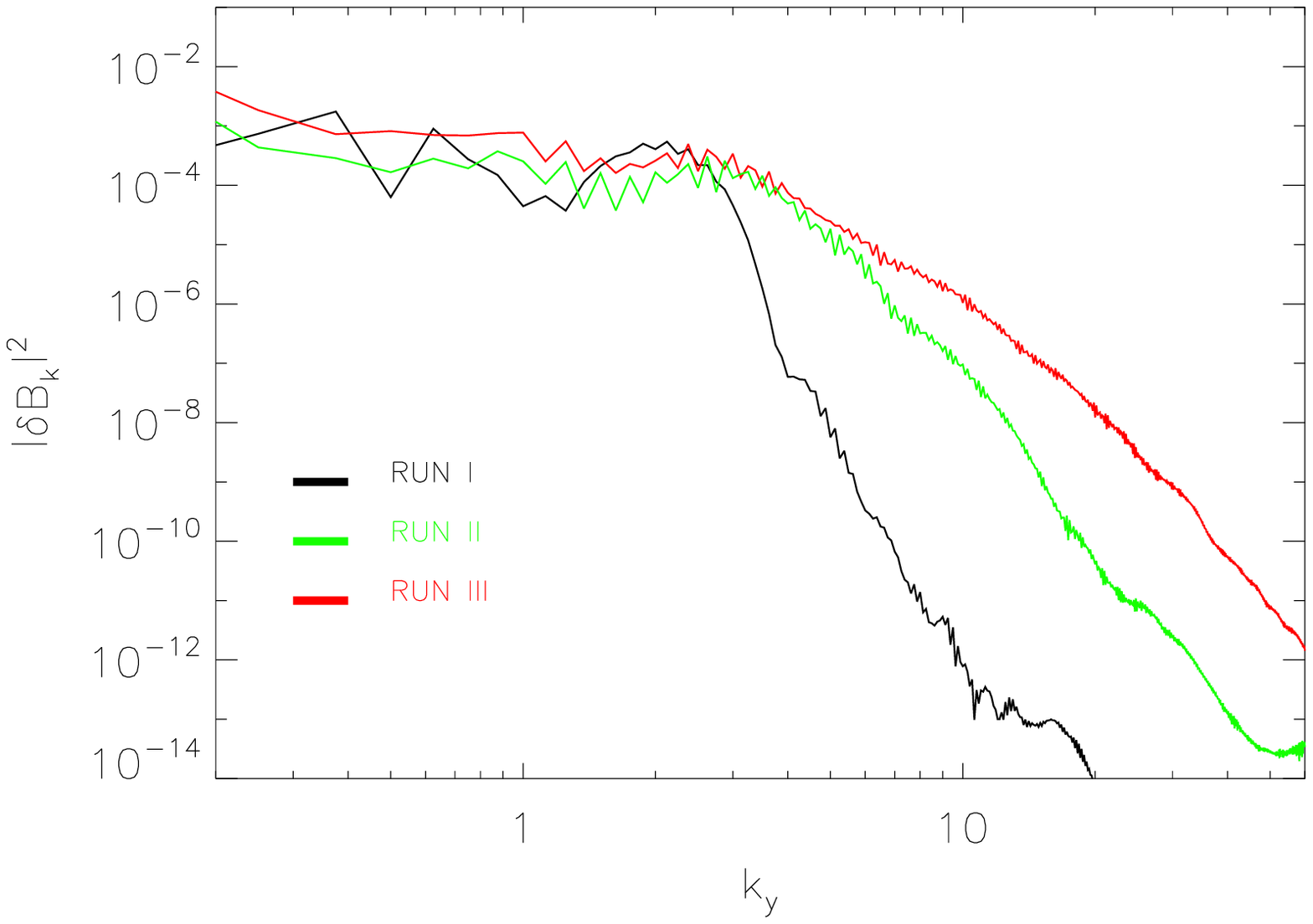}\\
 \includegraphics[width=0.55\textwidth]{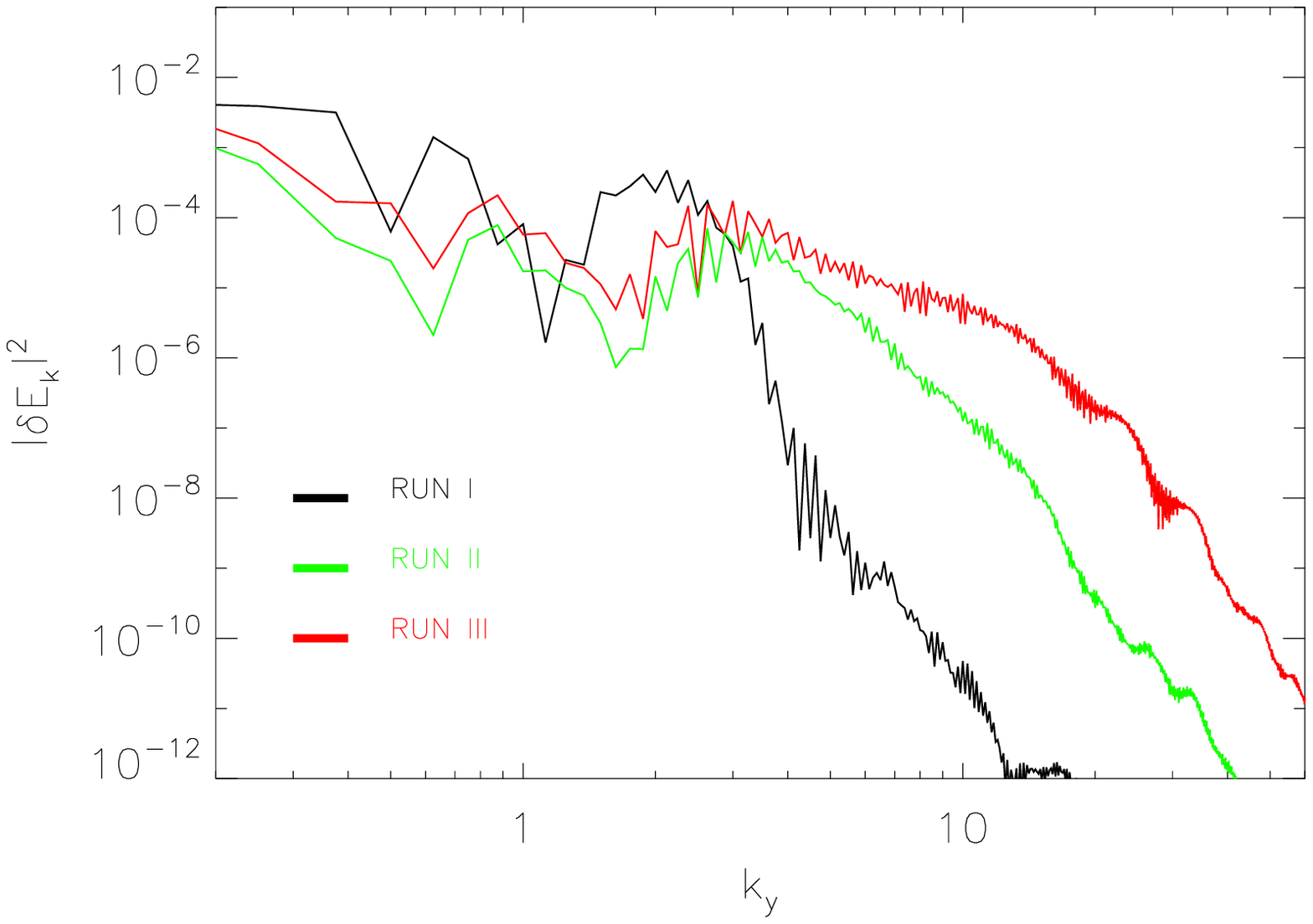}\\
 \includegraphics[width=0.55\textwidth]{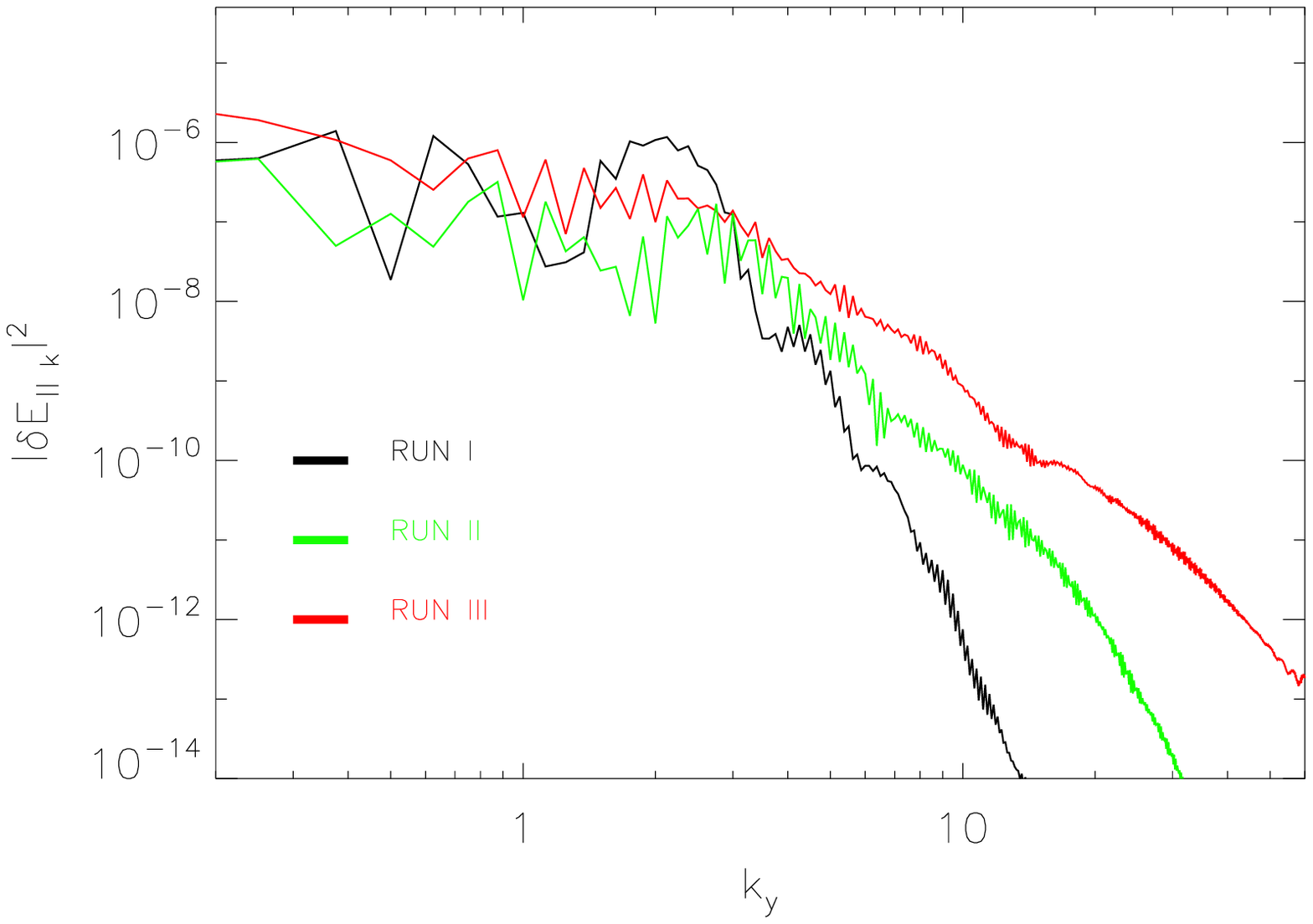}
 \caption{Power spectra of the total magnetic field (top panel), total electric field (middle panel) and electric field component
parallel to the local magnetic field (bottom panel), for RUN I (black line), RUN II (green line) and RUN III 
(red line).}
 \label{Fig:spettri}
\end{figure}

We also noticed that, when increasing the nonlinearity of the initial perturbation a clear transition to turbulence is recovered.
This can be appreciated in Figure \ref{Fig:spettri}, where the power spectra of magnetic $|\delta B_k|^2$ (top) and electric
$|\delta E_k|^2$ (middle) energy, summed over the parallel wavenumbers $k_x$ (reduced spectra), are plotted as a function of the
transverse wavenumber $k_y$. Here, for each field $g(x,y)$, we have computed $\delta g (x,y) = g(x,y) - \left<g(x,y) \right>_x$,
where $\left< \cdot \right>_x$ represents the mean value in the $x$-direction. It is clear from these plots that the magnetic and
electric energy content at wavenumbers larger than $k_yd_p\simeq 2$ is negligible for the case of RUN I (black curve), while
increases significantly for RUN II and RUN III (green and red curve, respectively). Moreover, at the bottom in the same figure we
display the power spectrum of the parallel electric energy $|\delta E_{\parallel k}|^2$. Here, the clear peak visible at
$k_yd_p\simeq 2$ in the case of RUN I (black curve) corresponds to the KAW fluctuations produced through phase mixing, as
discussed in details in \citep{vasconez15}. It is worth noticing that this peak disappears with increasing the spectrum extension
(green and red curves), meaning that the energy stored in KAW fluctuations cascades towards short spatial scales, when kinetic
processes come into play.
We point out that the final state reached in RUN III is not a state of fully developed turbulence, in the classical Kolmogorov
view, but it can be thought of as a state of increased nonlinearity, with respect to RUN I and RUN II; in fact, going
from RUN I to RUN III the signature of wave-like activity (the well defined bump visible in the spectra for RUN I) is gradually
lost and a significant amount of energy reaches the spectral range of high wavenumbers.

\begin{figure}
 \centering
 \includegraphics[width=0.45\textwidth]{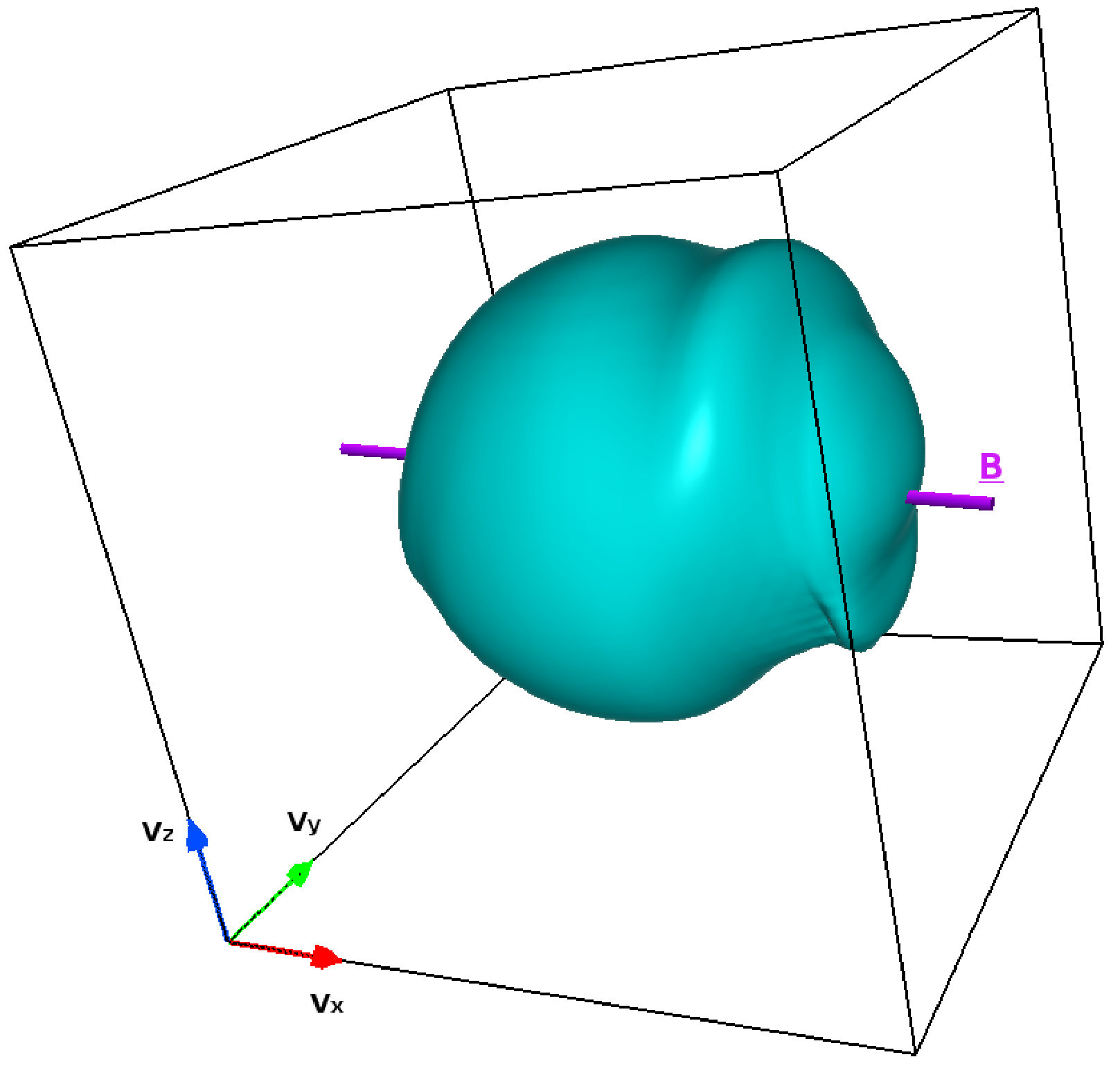}
 \includegraphics[width=0.45\textwidth]{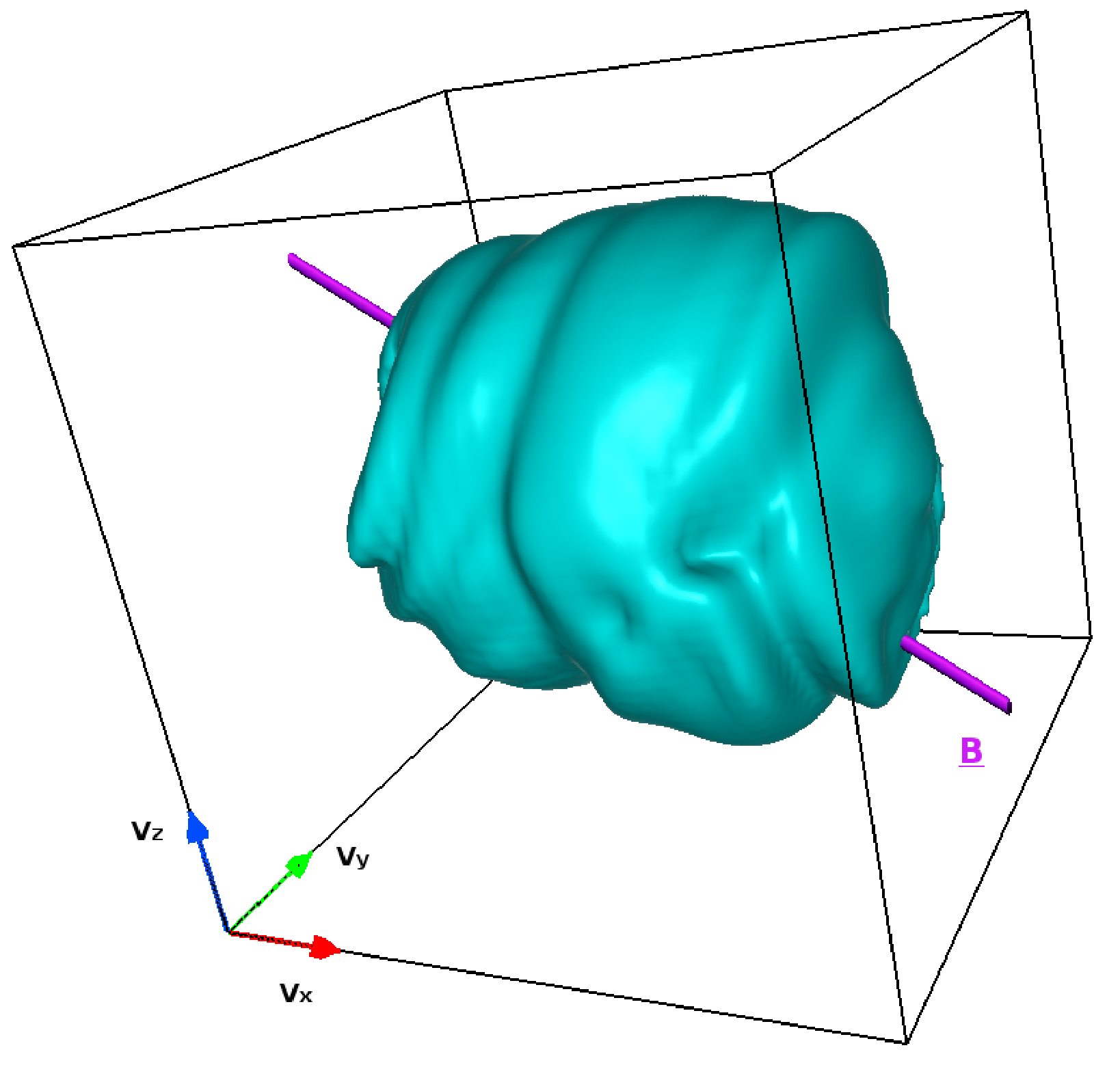}
 \includegraphics[width=0.45\textwidth]{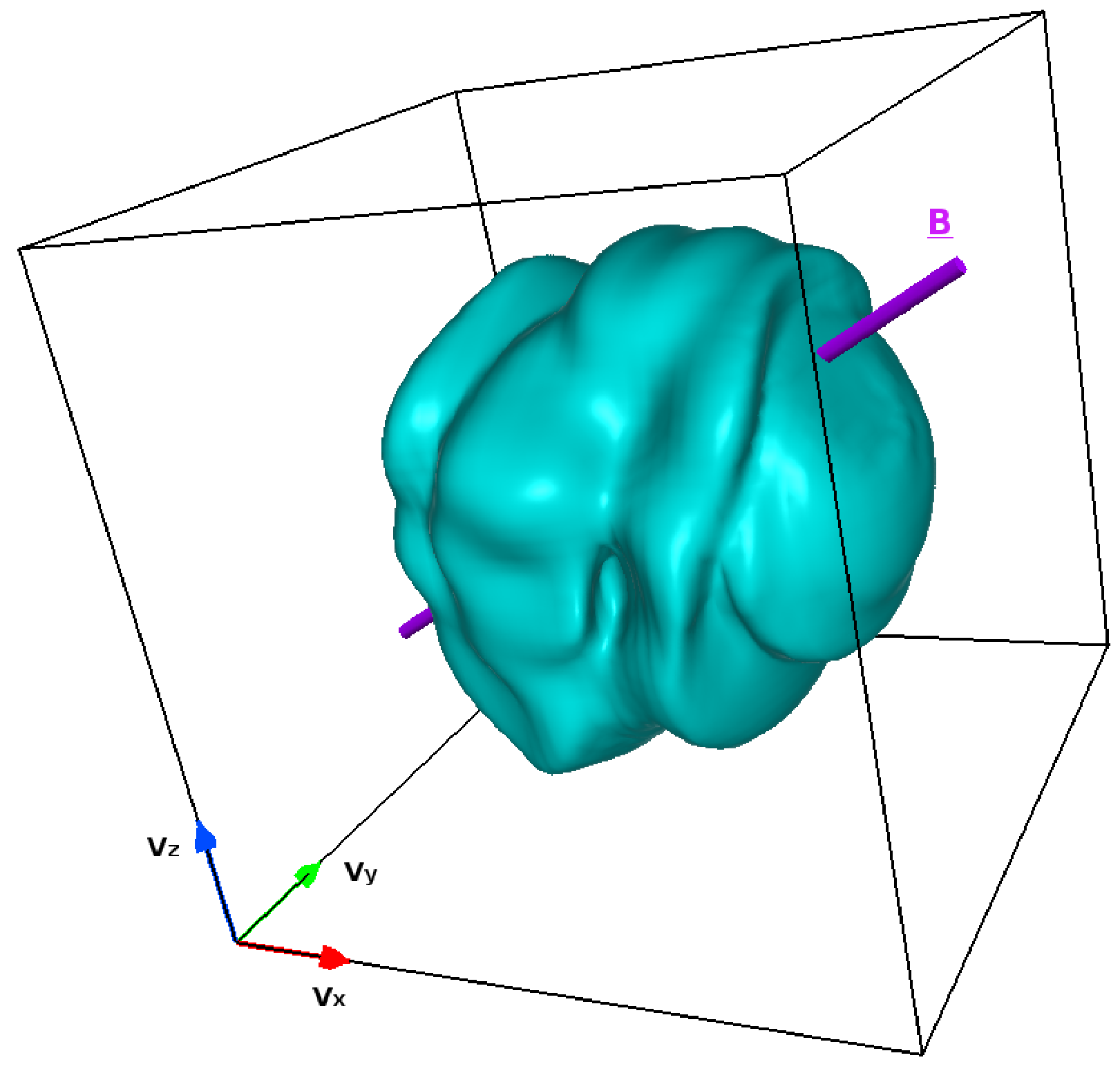}
 \caption{Iso-surface plot of the proton VDF in velocity space, at the spatial location where $\epsilon$ is maximum 
for RUN I (top left), RUN II (top right), and RUN III (bottom); the magenta tube in each plot indicates 
the direction of the local magnetic field.}
 \label{Fig:VDF}
\end{figure}

The Eulerian HVM code allows for an almost noise-free description of the proton distribution function in phase space, being, for
this reason, an indispensable tool to analyze the effects of kinetic processes on the plasma dynamics. In Figure \ref{Fig:VDF}, we
report the three-dimensional surface plots of the proton VDF at $t = 105$, computed at the spatial point where
$\epsilon=\epsilon_{max}$ for each run (these spatial points are located inside the shear regions, as can be seen in the middle
panels of Figure \ref{Fig:contours}). The unit vector of the local magnetic field is displayed in these plots as a magenta tube.
In the upper-left plot, corresponding to RUN I, one notices smooth deviations of the particle VDF from the spherical Maxwellian
shape, with the appearance of a barely visible bulge along the local field and a ring-like modulation in the perpendicular plane.
Here, the direction of the local field seems to be still a preferred direction of symmetry for the particle VDF. In RUN II
(upper-right plot) where nonlinearities are stronger, the particle VDF appears more distorted than in RUN I. Finally, in RUN III
(lower plot) where the transition to a turbulent state has been observed through the power spectra discussed above, any symmetry
of the VDF is lost, as sharp gradients and small-scale velocity structures have been produced through the nonlinear interaction of
protons with the fluctuating fields.

\begin{figure*}
 \centering
  \includegraphics[width=1.\textwidth,keepaspectratio]{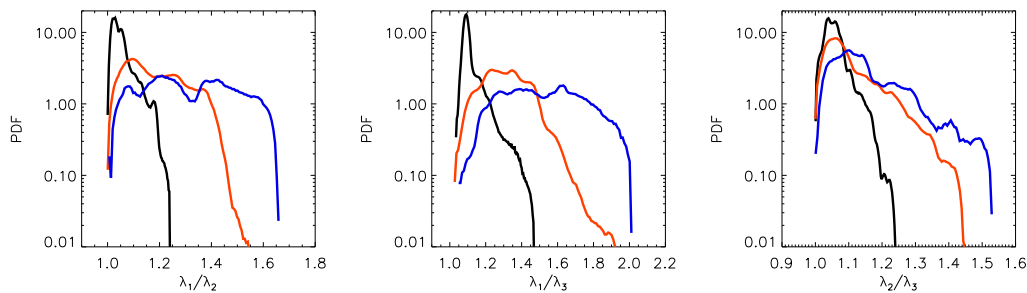}
 \caption{PDF of $\lambda_1/\lambda_2$ (left), $\lambda_1/\lambda_3$ (middle) and $\lambda_2/\lambda_3$ (right) from RUN III at
$t=105$, computed for three different ranges of values of $\epsilon$, namely $0\le\epsilon (t^\ast)\le\epsilon_{max}(t^\ast)/3$
(black curve), $\epsilon_{max}(t^\ast)/3<\epsilon (t^\ast)\le 2\epsilon_{max}(t^\ast)/3$ (red curve) and
$2\epsilon_{max}(t^\ast)/3\le\epsilon (t^\ast)\le \epsilon_{max}(t^\ast)$ (blue curve).}
 \label{Fig:PDF}
\end{figure*}

In order to provide a more quantitative description of the deviation of the VDFs from thermodynamic equilibrium, we computed the
preferred directions of $f$ in velocity space \citep{servidio12}, for each spatial position, from the stress tensor:
\begin{equation}
 \Pi_{i j}=n^{-1} \int (v_i - u_i)(v_j - u_j)f d^3 v,
 \label{pressure}
\end{equation}
This tensor can be diagonalized by computing its eigenvalues $\{\lambda_1,\lambda_2,\lambda_3\}$ (ordered in such way that
$\lambda_1 >\lambda_2 > \lambda_3$) and the corresponding normalized eigenvectors $\{ {\bf \hat{e}}_1, {\bf \hat{e}}_2,  {\bf
\hat{e}}_3\}$ which define the minimum variance frame (MVF). We point out that $\lambda_i$ are the temperatures and ${\bf
\hat{e}}_i$ the anisotropy directions of the VDF. The information given by the ratios $\lambda_i/\lambda_j$ is in some sense
included in $\epsilon$; nevertheless, the ratios of the eigenvalues evidently provide additional relevant insights into the
symmetry of VDF, which is important to investigate. Therefore, for RUN III at $t=t^*=105$, we computed the Probability
Distribution Function (PDF) of the ratios $\lambda_i/\lambda_j$ ($i,j=1,2,3$ and $j\ne i$), conditioned to the values of $\epsilon
(t=t^\ast)$. Note that each of these ratios is equal to unity for a Maxwellian VDF. In Figure \ref{Fig:PDF}, we show the PDF of
$\lambda_1/\lambda_2$ (left panel), $\lambda_1/\lambda_3$ (middle panel) and $\lambda_2/\lambda_3$ (right panel); these PDFs have
been computed for three different ranges of values of $\epsilon$, $0\le\epsilon (t^\ast)\le\epsilon_{max}(t^\ast)/3$ (black
curve), $\epsilon_{max}(t^\ast)/3<\epsilon (t^\ast)\le 2\epsilon_{max}(t^\ast)/3$ (red curve) and
$2\epsilon_{max}(t^\ast)/3\le\epsilon (t^\ast)\le \epsilon_{max}(t^\ast)$ (blue curve). It can be noticed from this figure that in
the range of small $\epsilon$ the three PDFs have a peak close to unity (they are not exactly centered around $1$, since the
minimum value of $\epsilon$ is not zero), this suggesting that, when the level of nonlinearity is low, the distribution
function can be slightly far from the Maxwellian shape, still keeping one (or more) axis of symmetry; on the other hand, as
$\epsilon$ increases, high tails appear in the PDF signals, this meaning that, in the case of significant deviations from
Maxwellian, is not possible to make assumptions on the shape of the VDF. As a consequence, the use of reduced models, based on
restrictive approximations on the symmetry of the VDF, is not appropriate and one must adopt more complete models able to describe
the evolution of the VDF in a full 3D velocity space. Note, in particular that, because of the multiple anisotropies observed, and
because of the misalignment with the ambient field, any gyrotropic approximation looses its validity, as we will further
demonstrate below.

\begin{figure}
 \centering
 \includegraphics[width=0.55\textwidth,keepaspectratio]{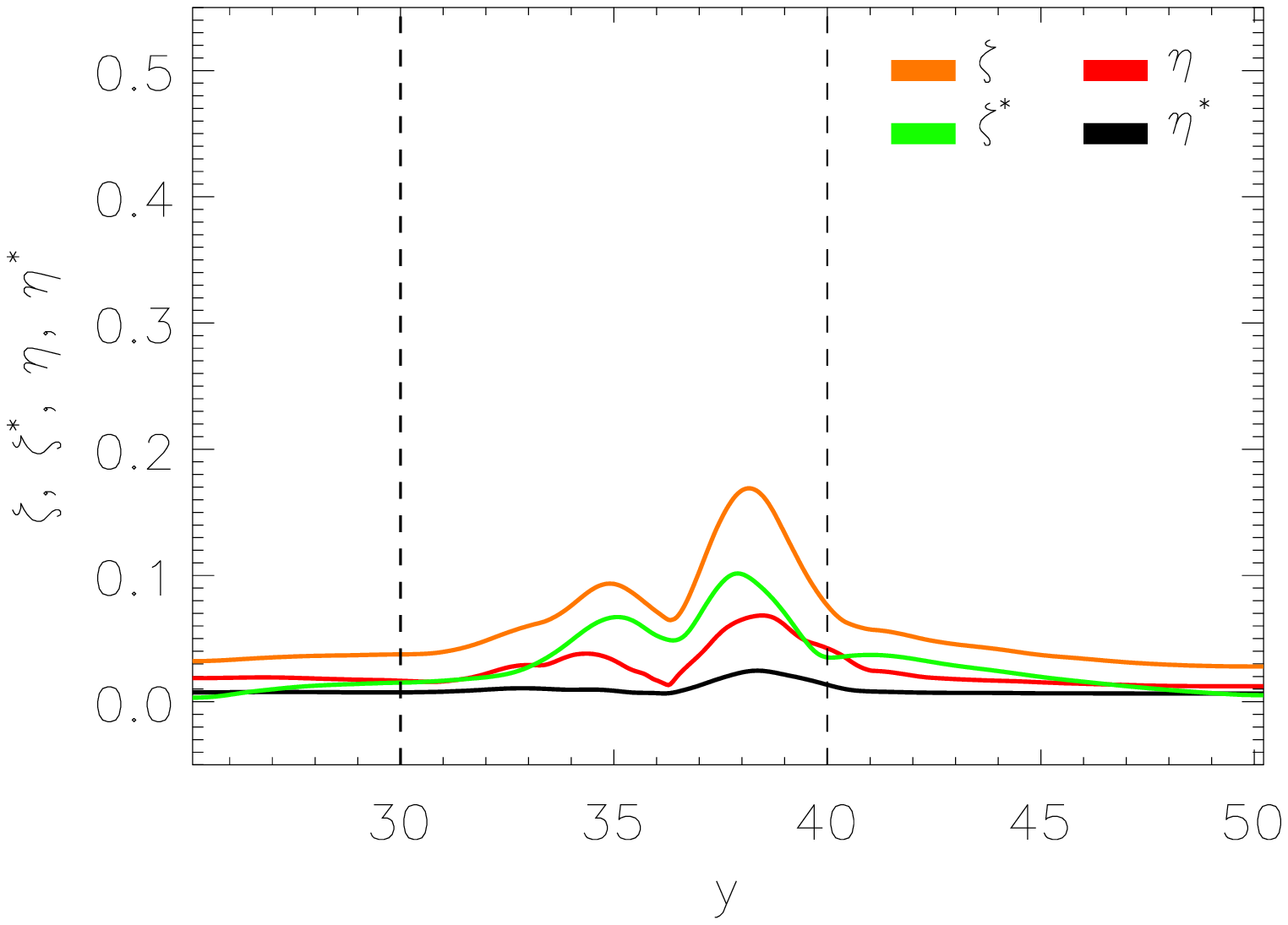}\\
 \includegraphics[width=0.55\textwidth,keepaspectratio]{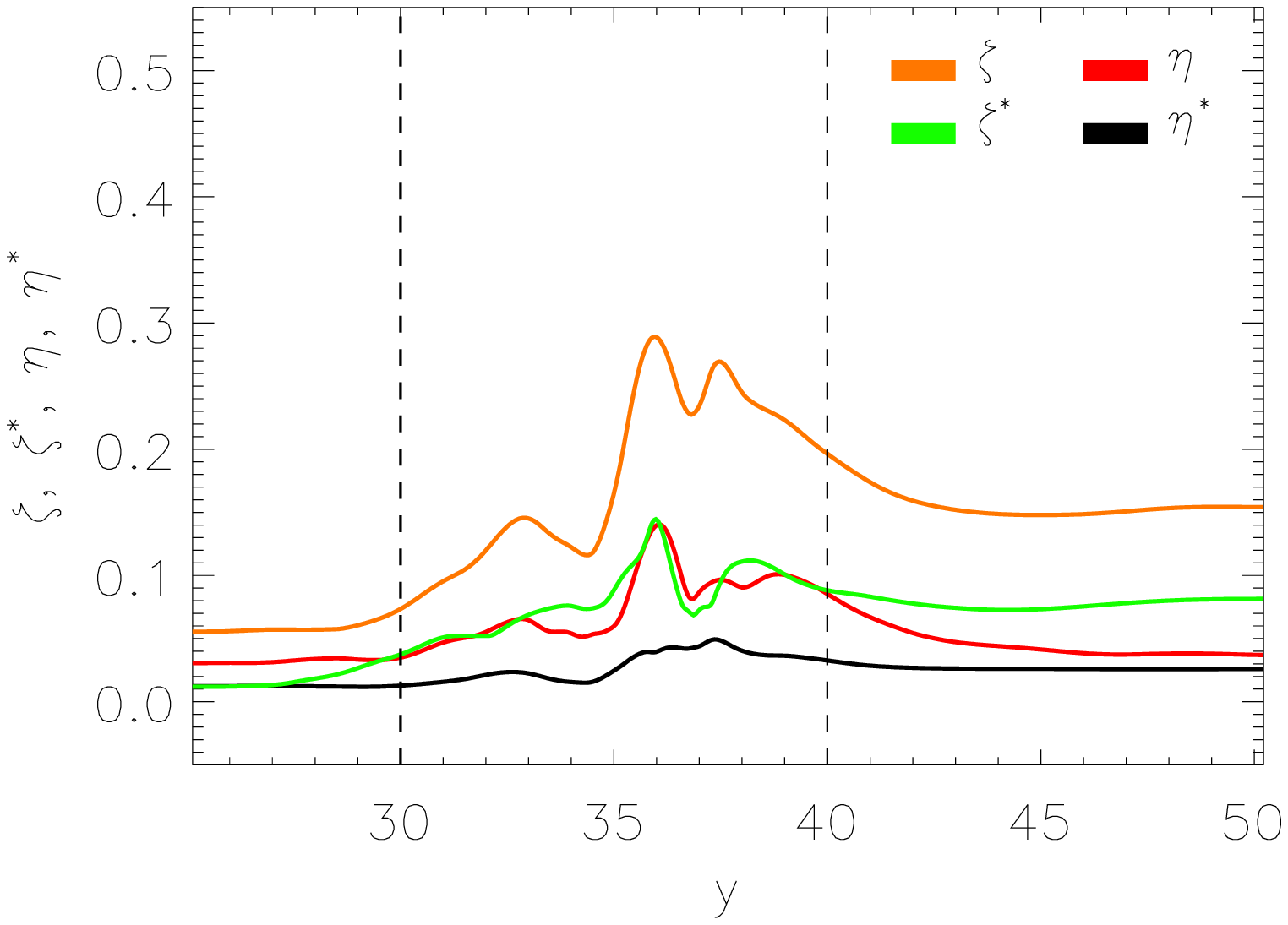}\\
 \includegraphics[width=0.55\textwidth,keepaspectratio]{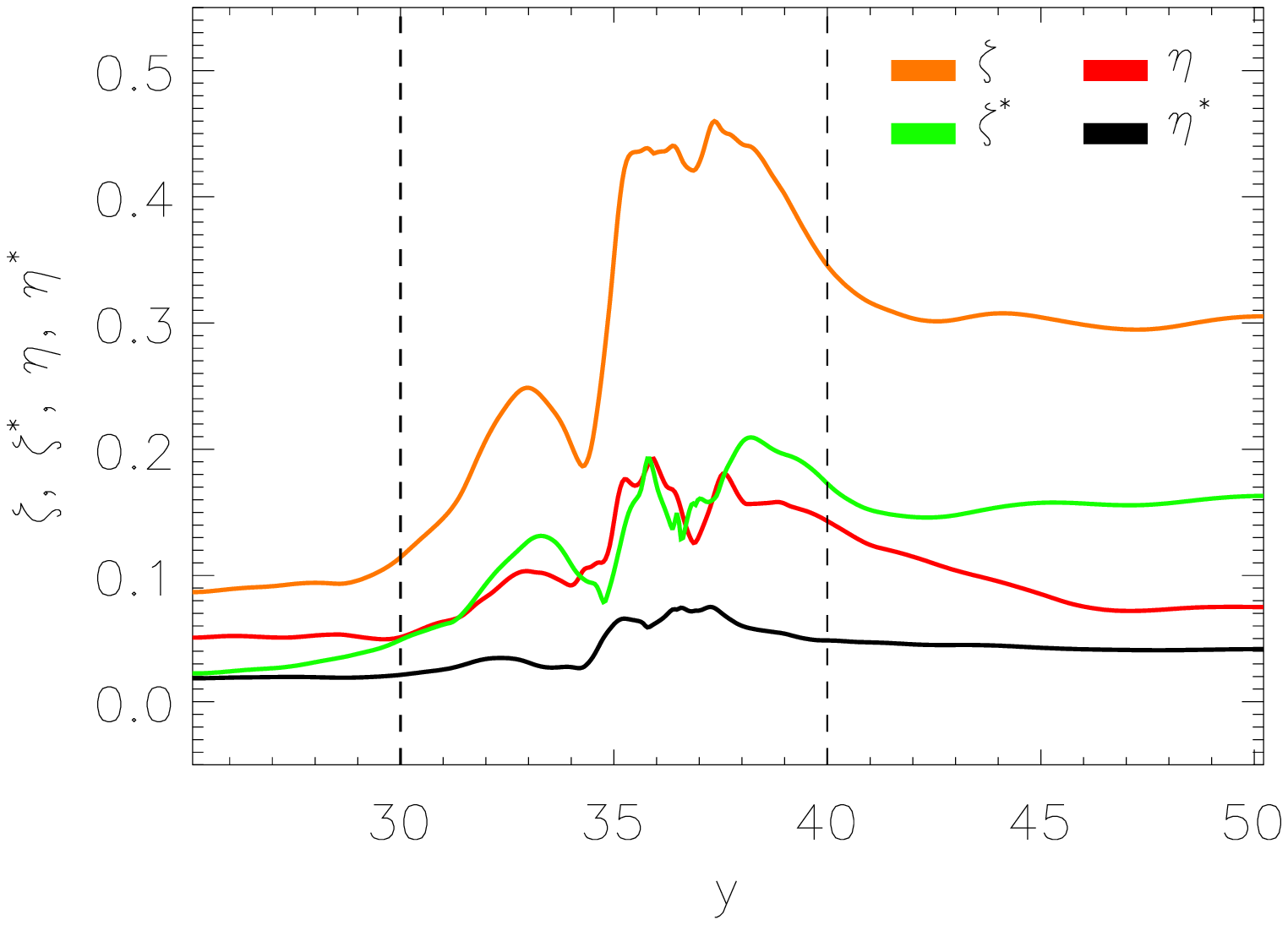}
 \caption{Anisotropy indexes $\zeta$ (orange line) and $\zeta^{\ast}$ (green line) and non-gyrotropy indexes $\eta$ (red line),
and $\eta^{\ast}$ (black line), averaged over $x$ and plotted as a function of $y$ in the interval $y=[L/2,L]$, for RUN I (top
panel), RUN II (middle panel) and RUN III (bottom panel).}
 \label{Fig:NG}
\end{figure}

With the aim of characterizing the nature of the deformation of the particle VDFs and to identify the spatial regions which are
the sites of kinetic activity, we computed two indexes of departure from Maxwellian, i. e. the temperature anisotropy index and
the gyrotropy index, in two different reference frames, namely the MVF and the local magnetic field frame (LMF). Therefore, we
define the anisotropy indicators $\zeta=|1-\lambda_1/\lambda_3|$ (MVF) and $\zeta^\ast=|1-T_\perp/T_\parallel|$ (LMF), where
$T_\perp$ and $T_\parallel$ are the temperatures with respect to the local magnetic field, and the gyrotropy indicator in the MVF
$\eta=|1-\lambda_2/\lambda_3|$. The gyrotropy indicator in the LMF $\eta^\ast$ can be computed by using the normalized Frobenius
norm of the nongyrotropic part $\Nv$ of the full pressure tensor ${\bf\Pi}$, introduced by \citet{aunai13}: 
\begin{equation}
 \eta^\ast = \frac{\sqrt{\sum_{i,j} N_{ij}^2}}{Tr({\bf\Pi})},
\end{equation}
\noindent
where $N_{ij}$ are the components of the tensor $\Nv$, and $Tr (\Nv) = 0$. It is worth to point out that all indexes defined above
are identically null if the particle VDF is Maxwellian. The contour plots of $\epsilon$ in the middle column of Figure
\ref{Fig:contours} show that the main departures from Maxwellian occur, for both runs, in the shear regions where current density
achieves its maximum value (though high values of $\epsilon$ are spread on a larger portion of the spatial domain around the
shear regions for RUN II and even more for RUN III). In Figure \ref{Fig:NG} we report the $y$ profile of the non-Maxwellianity
indexes $\zeta$ (anisotropy in the MVF), $\zeta*$ (anisotropy in the LMF), $\eta$ (non-gyrotropy in the MVF) and $\eta*$
(non-gyrotropy in the LMF), averaged over x for the three runs. We focus on the shear region on the right half of the
spatial simulation domain (delimited by vertical black-dashed lines). 
 
In the more turbulent situation of RUN III, nonlinear interaction of protons with large amplitude fluctuations of the KAW type
generates larger deviation of the particle VDF from Maxwellian in the shear region. Moreover, this effect is also visible in the
regions outside the shear; this could be due both to the transverse drift of KAWs from the shear regions towards the boundaries of
the numerical domain \citep{vasconez15}, and to nonlinear effects intrinsic to the initial perturbations which are stronger in
such lateral regions. On the other hand, for the quasi-linear RUN I, the particle VDF becomes anisotropic and non-gyrotropic as
viewed from both MVF and LMF, only in the shear regions. An intermediate situation is observed for RUN II.

In order to study quantitatively the generation of structures in velocity space due to kinetic processes, we implemented the
following procedure. We again focused on the most turbulent run (RUN III) at time $t=t^*=105$. We selected the spatial locations
where the maximum and the minimum values of $\epsilon$ are achieved at this time and considered the corresponding proton VDFs
$f_{max}({\bf v})$ and $f_{min}({\bf v})$ at these spatial locations. It is worth to note that $f_{max}({\bf v})$ is the VDF shown
in Figure \ref{Fig:VDF} (right). Therefore, we performed a rotation of $f_{max}({\bf v})$ and $f_{min}({\bf v})$, moving them into
the system ${\bf v'}$ in which the local magnetic field is along the $\hat{\bf e}_{v_z'}$ direction. At this point, we computed the 
quantities $\delta_{max}({\bf v'}) = f_{max}({\bf v'}) - f^{B}_{max}({\bf v'})$ and $\delta_{min}({\bf v'}) = f_{min}({\bf
v'}) - f^{B}_{min}({\bf v'})$, $f^{B}_{max/ min}$ being the bi-Maxwellian VDF evaluated through the velocity moments of
$f_{max/min}$. We remark that $\delta_{max}$ and $\delta_{min}$ store the information about the kinetic effects at work in the
system evolution, since, by subtracting from the VDFs their corresponding bi-Maxwellian, the main fluid-like effects have been 
ruled out. Note that there are several recent works where the attention has been concentrated on the dynamics of the
velocity space, very often invoked as entropy cascade \citep{tatsuno09,howes11,scheko16}. This new concept is very interesting
since it connects the cascade in physical space with a similar cascade in velocity space, where finer scales are formed and
finally dissipated through Landau damping or collisional mechanisms \citep{pezzi16}.

\begin{figure*}
 \centering
 \includegraphics[width=1\textwidth]{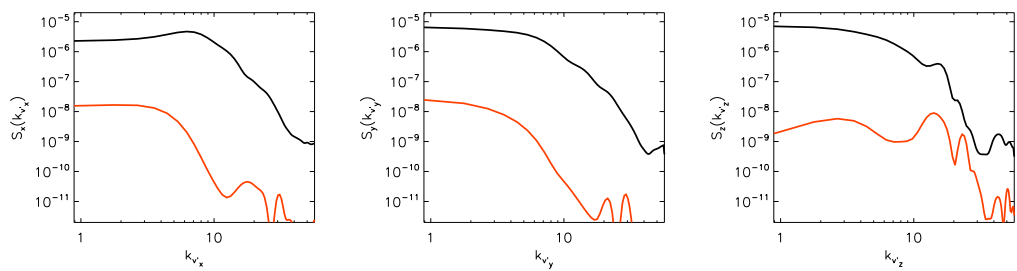}
 \caption{One-dimensional velocity space Fourier spectra $S_x(k_{v_x'})$ (left panel), $S_y(k_{v_y'})$ (middle panel) and
$S_z(k_{v_z'})$ (right panel) for $\delta f_{max}$ (black lines) and $\delta f_{min}$ (red lines), for RUN III at $t=105$.}
 \label{Fig:vspettri}
\end{figure*}

In order to analyze the velocity space structures, one can proceed in several ways. One possibility is given by the decomposition
of the VDF through Hermite polynomials \citep{tatsuno09,howes11,scheko16,pezzi16}. The advantage of this complete decomposition is
that one can capture all the features in velocity space (especially in our case, where high resolution VDFs are available). On the
other hand the interpretation of the eigenmodes in terms of classical scaling arguments (in analogy with the cascade in physical
space) may be quite complicated in a case where the VDF is fully three-dimensional and cannot be reduced to a simplified system,
by assuming, for example, gyrotropy. The second possibility is to decompose the fluctuations using classical Fourier
decomposition, in order to measure the intensity of the velocity space fluctuations, and understanding typical scales in velocity
space. Even though $\delta_{max}$ and $\delta_{min}$ are not formally periodic functions in velocity space, they go rapidly to
zero at the boundary of the simulation velocity box, allowing therefore the use of Fast Fourier transforms. For this reason, in
order to characterize the velocity space structures generated by the kinetic dynamics of protons, we performed a velocity space
Fourier transform of $\delta_{max}$ and $\delta_{min}$. In Figure \ref{Fig:vspettri}, we show the Fourier spectra
$S_x(k_{v_x'})$ (averaged over $k_{v_y'}$ and $k_{v_z'}$, left), $S_y(k_{v_y'})$ (averaged over $k_{v_x'}$ and $k_{v_z'}$, middle)
and $S_z(k_{v_z'})$ (averaged over $k_{v_x'}$ and $k_{v_y'}$, right), respectively, for $\delta_{max}$ (black line) and
$\delta_{min}$ (red line). Here, $k_{v_i'}$ represents the velocity space wave number associated with the velocity scale $v_i'$. 

One can easily realize that, in each direction, velocity space spectra of $\delta_{min}$ (VDF close to the Maxwellian shape) 
display a significantly lower power amplitude than those of $\delta_{max}$. Moreover, since the spectra of $\delta_{max}$ (black 
curves) have comparable energetic content in each direction, one may argue that there is no preferential direction for the VDF, 
when this is efficiently shaped by kinetic processes. In particular, the high $k_v$ bumps  may indicate the presence of
structures (beams and rings) in velocity space. An interesting analogy arises here with the cascade in physical space, in the
presence of strongly anisotropic turbulence. As it can be observed from Figure 8, for a low level of kinetic activity (red
curves), some isolated high-$k_v$ peaks can be observed, possibly related to local Cherenkov and/or cyclotron resonances
\citep{kennel66}. These are similar to the peaks observed due to wave activity in the Eulerian spectra \citep{dmitruk09}. On the
other hand, when the level of kinetic activity increases (black curves), these peaks disappear, and a more continuum cascade-like
spectrum is observed in velocity space, similarly to that recovered in physical space (cfr. with Fig. \ref{Fig:spettri}).
This is a preliminary study, which surely deserves a future investigation. This phenomenology could be related to the idea of the
“double cascade” (physical-velocity space), as suggested in several recent works \citep{tatsuno09,scheko16}.

\section{Summary and Conclusions}
As recently shown by \citet{vasconez15,pucci16}, Kinetic Alfv\'en waves are naturally generated through the phase mixing
mechanism, when Alfv\'en waves propagate in an inhomogeneous medium. In the present paper, we reproduced numerically, through
2D-3V Hybrid Vlasov-Maxwell simulations, the generation of KAWs by imposing Alfv\'enic perturbations on an initial pressure
balanced magnetic shear equilibrium. Both the characteristic of the initial equilibrium and the amplitude of the perturbations
have been varied, in order to explore the system dynamics in different regimes, focusing, in particular, on the transition from a
quasi-linear to a turbulent regime. Moreover, as the HVM code provides an almost noise-free description of the proton distribution
function, we have shown how the interaction of large amplitude KAW fluctuations with protons shapes the VDF and make it depart
from local thermodynamic equilibrium.

When comparing the quasi-linear RUN I with the more nonlinear RUN II and the turbulent RUN III, one realizes that many
interesting effects arise, as the amplitude of the perturbations increases and the gradients of the initial magnetic shear get
sharper. First of all, any wave-like activity, recognized in RUN I as well-defined bumps in the turbulent electric and magnetic
spectra, disappears in RUN II and RUN III, in which a significant amount of energy is recovered at short spatial scales. As a
consequence of this small-scale activity, the proton VDF results much more distorted in RUN III as compared to RUN I, loosing any
property of symmetry with respect to the direction of the local magnetic field. Indeed, as we discussed in detail in the previous
Section, in RUN III, in the spatial regions where strong departures from Maxwellian are observed, the proton VDF develops
temperature anisotropy, non-gyrotropy features and many other complicated deformations. 

We proposed to provide quantitative information on these kinetic distortions of the proton VDF, by employing different
non-Maxwellianity indexes, like the anisotropy index and the non-gyrotropy index, which have been computed both in the minimum
variance frame and in the local magnetic field frame. Interestingly, all these indexes behave in a similar way, achieving higher
values in the shear regions, where the initial magnetic configuration produced strong current sheets, while decreasing in the
homogeneous regions far from the shears. These results support the idea \citep{servidio12,valentini14} that kinetic effects are
not uniformly distributed in space, but are rather intermittent and localized in certain space regions determined by the topology
of the magnetic field.

Finally, through a Fourier analysis performed on the deviations of the proton VDF from a bi-Maxwellian, we pointed out that, when
kinetic effects are retained in the description of the plasma dynamics, beside a turbulent cascade in physical space, an 
analogous cascade is produced in velocity space \citep{tatsuno09,howes11,scheko16}, this emphasizing the physical link between the
small spatial scale structures driven by the turbulent cascade and the fine velocity gradients arising naturally through kinetic 
effects.

\section{Acknowledgements}
This work has been supported by the Agenzia Spaziale Italiana under the Contract No. ASI-INAF 2015-039-R.O “Missione M4 di ESA:
Partecipazione Italiana alla fase di assessment della missione THOR.” The numerical HVM simulations have been run on the Fermi
supercomputer at CINECA (Bologna, Italy), within the ISCRA-C projects IsC26-COLTURBO, IsC26-PMKAW and IsC25-MHDwMoPC.


\end{document}